\documentclass[ %
 twocolumn, 
 10pt,
nofootinbib,
 amsmath,amssymb,
 aps,
 prd,
 floatfix,
 superscriptaddress
]{revtex4-2}

\usepackage{graphicx}
\usepackage{dcolumn}
\usepackage{bm}
\usepackage{rotating} 
\usepackage{diagbox}

\usepackage{lineno}
\usepackage{acro}
\usepackage{makecell}
\usepackage{multirow}
\usepackage{xcolor}
\usepackage[colorlinks,linktocpage,linkcolor=cyan,citecolor=cyan]{hyperref}

\DeclareAcronym{DESI}{
  short = DESI ,
  long = Dark Energy Spectroscopic Instrument ,
  short-plural =  ,
}
\DeclareAcronym{DR2}{
  short = DR2 ,
  long = Data Release 2 ,
  short-plural =  ,
}
\DeclareAcronym{DR6}{
  short = DR6 ,
  long = Data Release 6 ,
  short-plural =  ,
}
\DeclareAcronym{CMB}{
  short = CMB ,
  long = cosmic microwave background ,
  short-plural =  ,
}
\DeclareAcronym{BAO}{
  short = BAO ,
  long = baryon acoustic oscillation ,
  short-plural =  ,
}
\DeclareAcronym{BBN}{
  short = BBN ,
  long = Big Bang nucleosynthesis ,
  short-plural =  ,
}
\DeclareAcronym{DH}{
  short = DH ,
  long = degenerate hierarchy ,
  short-plural =  ,
}
\DeclareAcronym{NH}{
  short = NH ,
  long = normal hierarchy ,
  short-plural =  ,
}
\DeclareAcronym{IH}{
  short = IH ,
  long = inverted hierarchy ,
  short-plural =  ,
}
\DeclareAcronym{CL}{
  short = CL ,
  long = confidence level ,
  short-plural =  ,
}
\DeclareAcronym{NDI}{
  short = NDI ,
  long = neutrino density isocurvature ,
  short-plural =  ,
}
\DeclareAcronym{ACT}{
  short = ACT ,
  long = Atacama Cosmology Telescope ,
  short-plural =  ,
}
\DeclareAcronym{SPT}{
  short = SPT ,
  long = South Pole Telescope ,
  short-plural =  ,
}
\DeclareAcronym{SPTthreeG}{
  short = SPT-3G ,
  long = South Pole Telescope third-generation ,
  short-plural =  ,
}
\DeclareAcronym{DES}{
  short = DES ,
  long = Dark Energy Survey ,
  short-plural =  ,
}
\DeclareAcronym{SN}{
  short = SN ,
  long = supernova ,
  short-plural = e ,
}
\DeclareAcronym{CPL}{
  short = CPL ,
  long = Chevallier–Polarski–Linder ,
  short-plural =  ,
}
\DeclareAcronym{MCMC}{
  short = MCMC ,
  long = Markov chain Monte Carlo ,
  short-plural =  ,
}
\DeclareAcronym{CLASS}{
  short = CLASS ,
  long = Cosmic Linear Anisotropy Solving System ,
  short-plural =  ,
}

\begin{document}

\title{Constraints on the Sum of Neutrino Masses from ACT DR6 and DESI DR2 Considering Isocurvature Initial Conditions}

\author{Hongsheng Hou}
\affiliation{School of Physics, Hangzhou Normal University, No.2318 Yuhangtang Road, Yuhang District, Hangzhou 311121, China}

\author{Sai Wang}
\thanks{Corresponding author}
\email{wangsai@hznu.edu.cn}
\affiliation{School of Physics, Hangzhou Normal University, No.2318 Yuhangtang Road, Yuhang District, Hangzhou 311121, China}

\author{Zhi-Chao Zhao}
\thanks{Corresponding author}
\email{zhaozc@cau.edu.cn}
\affiliation{Department of Applied Physics, College of Science, China Agricultural University, 17 Qinghua East Road, Haidian District, Beijing 100083, China}

\author{Xin Zhang}
\thanks{Corresponding author}
\email{zhangxin@mail.neu.edu.cn}
\affiliation{Liaoning Key Laboratory of Cosmology and Astrophysics, College of Sciences, Northeastern University, Shenyang 110819, China}
\affiliation{National Frontiers Science Center for Industrial Intelligence and Systems Optimization, Northeastern University, Shenyang 110819, China}
\affiliation{MOE Key Laboratory of Data Analytics and Optimization for Smart Industry, Northeastern University, Shenyang 110819, China}

\begin{abstract}
We present a robust assessment of cosmological constraints on the sum of neutrino masses ($\sum m_\nu$) when relaxing the standard assumption of purely adiabatic primordial initial conditions. Allowing for a neutrino density isocurvature (NDI) component alongside the adiabatic mode, we analyse the latest CMB-SPA combination (Planck 2018, ACT DR6, and SPT-3G), DESI DR2 baryon acoustic oscillation data, and the DES Year 5 supernova sample. Within the $\Lambda$CDM model, the 95\% upper limit weakens only marginally from $\sum m_\nu < 0.052$ eV (purely adiabatic) to $< 0.057$ eV (including NDI), with the NDI amplitude consistent with zero. In the CPL dynamical dark energy model, the adiabatic limit is $< 0.111$ eV, shifting to $< 0.115$ eV with NDI, yet the isocurvature mode remains undetected. While these limits are robust against the inclusion of isocurvature perturbations, they are highly sensitive to both the assumed dark energy equation of state and the prior lower bound on $\sum m_\nu$. Notably, the adiabatic $\Lambda$CDM limit of $0.052$ eV lies below the minimum sum required by the normal neutrino mass hierarchy ($0.05878$ eV), indicating that this bound is an artifact of the statistical prior extending to zero. Imposing a physically motivated hierarchy-informed prior raises the limit to $< 0.092$ eV. Our results demonstrate that current data show no evidence for NDI modes and that the inferred neutrino mass upper limit is robust against this extension, but a definitive, model-independent bound requires addressing prior dependencies and dark energy uncertainties. This work provides the first joint constraint on $\sum m_\nu$ and NDI using the full CMB-SPA+DESI DR2+DES dataset.
\end{abstract}

\maketitle

\section{Introduction}\label{sec:introduction}

Understanding the properties of neutrinos, particularly their absolute mass scale and mass hierarchy, remains a fundamental challenge in modern physics. Neutrino oscillation experiments have firmly established two independent mass-squared differences: $\Delta m^2_{21} \approx 7.5\times10^{-5}~\text{eV}^2$ (solar splitting) and $|\Delta m^2_{31}| \approx 2.5\times10^{-3}~\text{eV}^2$ (atmospheric splitting) \cite{Esteban:2024eli}. However, these measurements only constrain the differences between squared masses, leaving the absolute scale and the ordering (normal vs. inverted) undetermined \cite{deSalas:2020pgw}. Laboratory searches for the effective electron neutrino mass, such as KATRIN, set an upper limit $m_\nu<0.45$~eV at $90\%$ \ac{CL} \cite{KATRIN:2024cdt}, while neutrinoless double-beta decay experiments like KamLAND-Zen provide upper limits on the effective Majorana mass in the range $28$--$122$~meV \cite{KamLAND-Zen:2024eml}. Despite these achievements, terrestrial bounds remain relatively weak and model-dependent \cite{Wong:2011ip}. Consequently, the sum of neutrino masses $\sum m_\nu$ is poorly constrained by laboratory experiments alone, motivating the use of complementary cosmological probes \cite{Lesgourgues:2006nd,Wong:2011ip,Zhao:2025evq,Zhang:2015uhk,Zhao:2016ecj,Zhang:2017rbg,Vagnozzi:2018jhn,RoyChoudhury:2018gay,Du:2024pai,Guo:2017hea,Huang:2015wrx,Wang:2017htc,Wang:2016tsz,Du:2025xes}.

Cosmological observations provide powerful and independent constraints on $\sum m_\nu$, often yielding significantly tighter upper limits \cite{Lesgourgues:2006nd, Wong:2011ip}. Massive neutrinos suppress the growth of matter density fluctuations on scales smaller than their free-streaming length, leaving distinct imprints on the \ac{CMB} and the large-scale structure of the Universe \cite{Lesgourgues:2006nd, Wong:2011ip}. Planck 2018 alone gives $\sum m_\nu<0.24$~eV, and combining with \ac{BAO} data tightens the limit considerably \cite{Planck:2018vyg}. Recently, the \ac{DESI} \ac{DR2} has provided highly accurate \ac{BAO} measurements over a wide redshift range, breaking degeneracies between $\sum m_\nu$ and other cosmological parameters such as $\Omega_m$ and $H_0$ \cite{DESI:2025zgx, Elbers:2025vlz}. Moreover, the \ac{ACT} \ac{DR6} has delivered high-resolution \ac{CMB} measurements that further sharpen these constraints, with recent joint \ac{ACT}, \ac{SPTthreeG}, and Planck analyses finding sub-$0.1$~eV upper limits on $\sum m_\nu$ at $95\%$ \ac{CL} within the $\Lambda$CDM framework \cite{ACT:2023kun, AtacamaCosmologyTelescope:2025blo, SPT-3G:2024atg, SPT-3G:2025bzu}. These cosmological bounds are already beginning to disfavor the inverted hierarchy, which requires $\sum m_\nu\gtrsim0.09892$~eV \cite{Esteban:2024eli, deSalas:2020pgw}.

However, cosmological constraints on $\sum m_\nu$ are inherently model-dependent, because they rely on specific assumptions about the primordial initial conditions \cite{Planck:2018vyg, Buckley:2025zgh}. The standard $\Lambda$CDM model assumes purely adiabatic initial conditions, where all species share the same curvature perturbation \cite{Planck:2018vyg, Bucher:2000kb}. This assumption is well-motivated by the simplest inflationary scenarios, but it is not inevitable \cite{Langlois:1999dw, Gordon:2000hv}. Neutrinos, being weakly interacting and massive, could in principle support an entirely different mode: the  \ac{NDI} mode, in which the photon-to-neutrino density ratio varies spatially while the total density perturbation vanishes initially \cite{Bucher:2000kb}. More generally, mixed adiabatic and isocurvature initial conditions arise naturally in many extensions of inflation, such as those involving multiple scalar fields or non-standard reheating \cite{Langlois:1999dw, Gordon:2000hv}. The presence of  \ac{NDI} modes would alter the \ac{CMB} power spectrum in ways distinct from the effect of neutrino mass, and future experiments could significantly reduce isocurvature error bars \cite{Buckley:2025zgh, Euclid:2025hlc}. Therefore, to robustly interpret current and future data, it is essential to relax the pure adiabatic assumption and explore how the $\sum m_\nu$ constraints vary when  \ac{NDI} is allowed.

The recent \ac{ACT}, \ac{SPTthreeG}, and \ac{DESI} observations offer significant advantages over previous-generation experiments, enabling much tighter constraints on $\sum m_\nu$. In the \ac{CMB} sector, Planck lacked the angular resolution to fully resolve small-scale structures \cite{Planck:2018vyg, ACT:2023kun}. The current generation of ground-based experiments, including \ac{ACT} and \ac{SPTthreeG}, have overcome this limitation \cite{ACT:2023kun, SPT-3G:2025bzu}. \ac{ACT} \ac{DR6} delivers high-resolution temperature and polarization maps over a wide sky area ($\sim23\%$), with arcminute-scale resolution and low noise levels, particularly in polarization \cite{AtacamaCosmologyTelescope:2025blo, ACT:2023kun}. Its combination with Planck is highly complementary: Planck fixes the large-scale baseline, while \ac{ACT} sharpens constraints on the damping tail and lensing, reducing degeneracies between $\sum m_\nu$ and parameters such as $\Omega_m$ and $\sigma_8$ \cite{AtacamaCosmologyTelescope:2025nti}. In parallel, \ac{SPTthreeG}, the other leading third-generation ground-based \ac{CMB} experiment, focuses on a deep $3.5\%$ field at the South Pole, achieving exceptional polarization sensitivity \cite{SPT-3G:2024atg, SPT-3G:2025bzu}. Its two-year delensed EE and lensing data provide an independent high-precision measurement of small-scale \ac{CMB} polarization, offering a crucial cross-check for systematic effects \cite{SPT-3G:2024atg, SPT-3G:2025bzu}. Together, \ac{ACT} and \ac{SPTthreeG} form the core of our CMB-SPA likelihood, covering complementary sky regions and providing joint sensitivity to both wide-angle and deep-polarization information \cite{SPT-3G:2025bzu}. For the large-scale-structure sector, DESI DR2 represents a major advance. Based on the first three years of DESI observations, the DR2 galaxy and quasar BAO analysis uses over 14 million discrete tracers, while the companion Ly$\alpha$-forest analysis achieves a combined $0.65\%$ precision on the isotropic BAO scale at $z_{\rm eff}=2.33$~\citep{DESI:2025zgx,DESI:2025zpo}. Consequently, the enhanced small-scale \ac{CMB} information from \ac{ACT} and \ac{SPTthreeG}, together with the high-precision \ac{BAO} data from \ac{DESI}, systematically tightens the limits on $\sum m_\nu$ and provides a crucial benchmark for future neutrino mass measurements \cite{DESI:2025zgx, SPT-3G:2025bzu}.

In this paper, we take full advantage of these state-of-the-art data sets to systematically investigate the robustness of cosmological upper limits on $\sum m_\nu$ against the assumption of initial conditions. We analyze the high-resolution \ac{CMB} measurements from \ac{ACT} \ac{DR6} and \ac{SPTthreeG}, the Planck 2018 \ac{CMB} data, the \ac{BAO} measurements from \ac{DESI} \ac{DR2}, and the Type Ia \ac{SN} sample from the \ac{DES} Year 5 \cite{DES:2024jxu, DES:2025sig}. By comparing constraints derived under the standard adiabatic scenario with those obtained when allowing for  \ac{NDI} initial conditions, we assess whether the bounds on $\sum m_\nu$ remain stable or become significantly relaxed. Our findings will demonstrate the degree to which current cosmological constraints on neutrino masses are model-dependent, or conversely, whether they exhibit strong robustness against extensions to the initial condition sector. This work provides crucial guidance for interpreting cosmological neutrino mass limits in the context of particle physics and for planning future surveys aimed at uncovering the absolute neutrino mass scale and its hierarchy.

The remainder of this paper is organized as follows. Section~\ref{sec:theory} describes the theoretical framework for massive neutrinos and  \ac{NDI} initial conditions, including their impact on the \ac{CMB} and matter power spectra. Section~\ref{sec:data_method} presents the observational data sets and the methodological details of our parameter inference. Section~\ref{sec:results} reports our main results, including constraints on $\sum m_\nu$ under adiabatic and isocurvature initial conditions. Finally, Section~\ref{sec:conclusion} discusses the robustness of the neutrino mass bound and provides concluding remarks.

\section{Theory}\label{sec:theory}

\begin{figure*}[htbp]
    \centering
    \includegraphics[width=.9\linewidth]{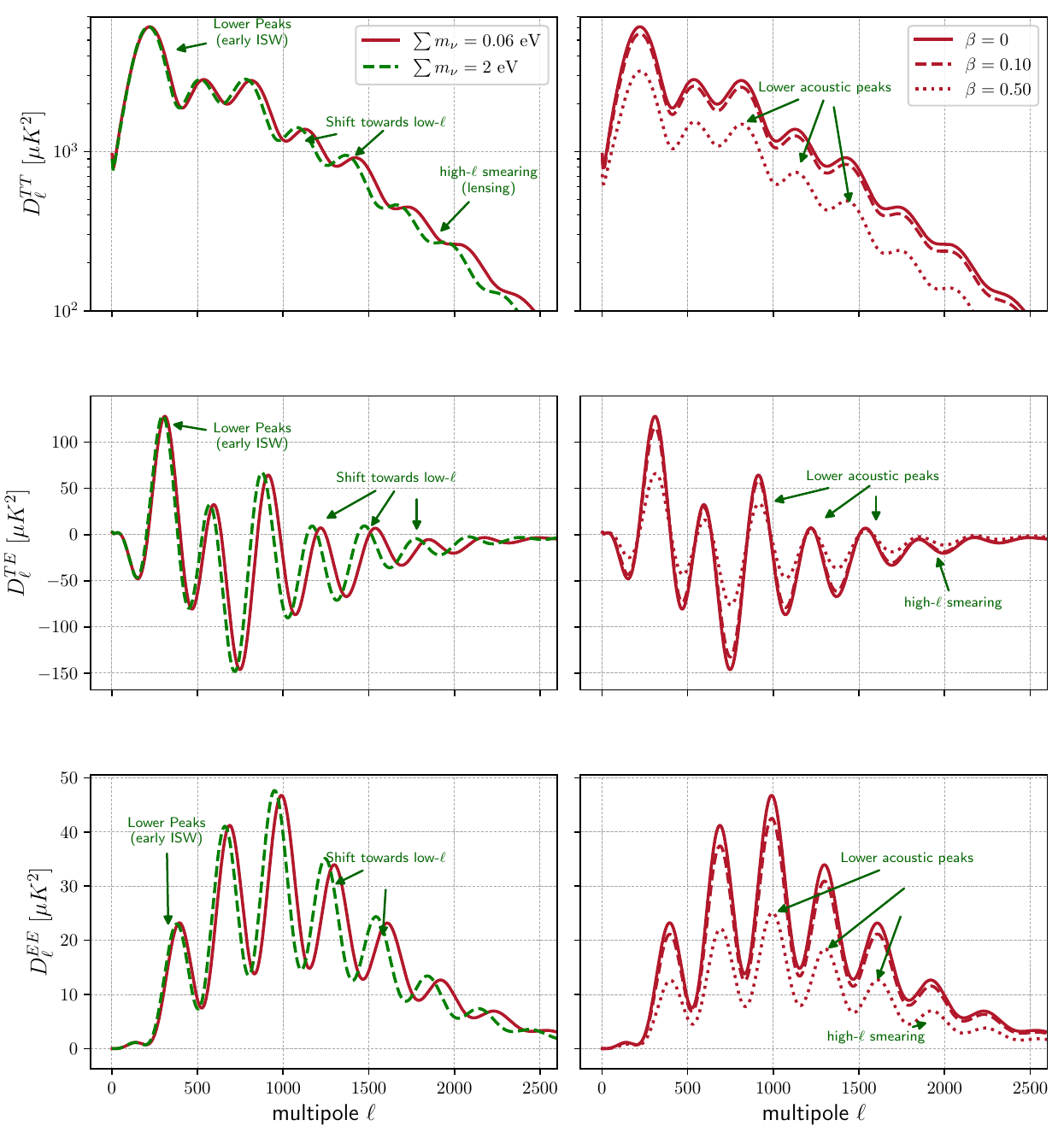}
    \caption{Lensed CMB temperature and polarization angular power spectra assuming degenerate massive neutrinos. In the left panels, the predictions for a fixed  \ac{NDI} fraction $\beta=0$ and varying sums of neutrino masses $\sum m_{\nu}=0.06\,\mathrm{eV}$ and $2\,\mathrm{eV}$ are compared. In the right panels, the sum of neutrino masses is fixed at $\sum m_{\nu}=0.06\,\mathrm{eV}$ while the  \ac{NDI} fraction is varied over $\beta=0$, $0.10$, and $0.50$, where $\beta(k_1)=\beta(k_2)\equiv\beta$ is assumed here. Colors identify the neutrino mass, while line styles identify the  \ac{NDI} fraction.}
    \label{fig:cmb_ndi}
\end{figure*}

\begin{figure*}[htbp]
    \centering
    \includegraphics[width=.9\linewidth]{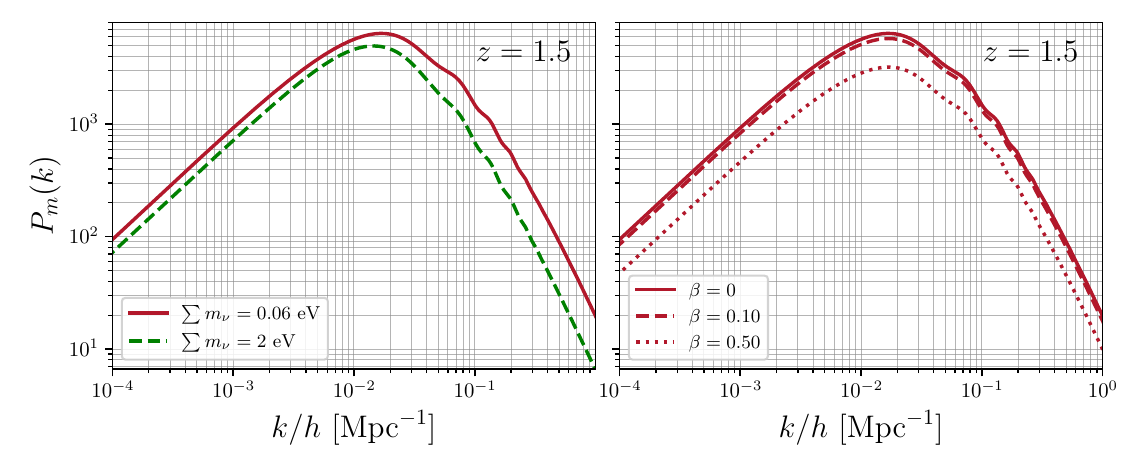}
    \caption{The same as Fig.~\ref{fig:cmb_ndi}, but the matter power spectrum $P_m(k)$ at $z=1.5$ is shown here. }
    \label{fig:bao_ndi}
\end{figure*}

Massive neutrinos affect cosmological observables primarily through their free-streaming behavior. In the standard $\Lambda$CDM model with adiabatic initial conditions, neutrinos with mass sum $\sum m_\nu$ suppress the growth of matter density perturbations on scales smaller than their free-streaming length, reducing the matter power spectrum and damping the small-scale \ac{CMB} anisotropies, particularly at high multipoles $\ell$ where the Silk damping tail is modified \cite{Lesgourgues:2006nd, Wong:2011ip}. See left panels of Figure~\ref{fig:cmb_ndi}. This suppression becomes more pronounced as $\sum m_\nu$ increases \cite{Lesgourgues:2006nd}. However, the effect is degenerate with other parameters such as the matter density $\Omega_m$ and the Hubble constant $H_0$ \cite{Planck:2018vyg, DESI:2025zgx}. These degeneracies can be broken by combining \ac{CMB} data with \ac{BAO} measurements, which provide a precise standard ruler at low redshifts \cite{Planck:2018vyg, DESI:2025zgx}. \ac{BAO} observables, specifically the angular diameter distance $D_A(z)$ and the Hubble expansion rate $H(z)$, are sensitive to the overall matter density and the expansion history, thereby helping to isolate the signature of neutrino mass \cite{DESI:2025zgx, DESI:2025zpo}. See left panels of Figure~\ref{fig:bao_ndi}. Hence, joint analyses of \ac{CMB} and \ac{BAO} data yield tight upper limits on $\sum m_\nu$ under the assumption of purely adiabatic initial conditions \cite{Planck:2018vyg, AtacamaCosmologyTelescope:2025blo, SPT-3G:2025bzu}.

In addition to the adiabatic mode, primordial perturbations may contain an  \ac{NDI} component, where the photon-to-neutrino density ratio varies spatially while the total density perturbation vanishes initially \cite{Bucher:2000kb}. The  \ac{NDI} mode modifies the \ac{CMB} power spectrum in ways distinct from the effect of neutrino mass. Unlike the adiabatic mode, which produces a pronounced Sachs--Wolfe plateau at low multipoles, the  \ac{NDI} contribution rises from $\ell \sim 10$, peaks around $\ell \sim 100$, and then declines, thereby suppressing the large-scale \ac{CMB} anisotropy and altering the relative heights of the acoustic peaks \cite{Bucher:2000kb, Buckley:2025zgh}. See right panels of Figure~\ref{fig:cmb_ndi}. In the matter power spectrum, the  \ac{NDI} component causes a scale-dependent suppression on intermediate wavenumbers ($k \sim 0.01$--$0.1\,h\,\text{Mpc}^{-1}$) because the gravitational potential decays more rapidly after horizon entry in the absence of an initial curvature perturbation \cite{Buckley:2025zgh}. See right panels of Figure~\ref{fig:bao_ndi}. For \ac{BAO} observables, the  \ac{NDI} component shifts the sound horizon at recombination $r_s$ and modifies the late-time distance-redshift relation $D_A(z)$, leading to a different redshift dependence of the \ac{BAO} angular scale $\theta_{\rm BAO}(z)=r_s/D_A(z)$ compared to the pure adiabatic case \cite{Buckley:2025zgh}. In this work, we consider a mixed scenario where an  \ac{NDI} component is present alongside the dominant adiabatic mode, and we investigate how the inclusion of  \ac{NDI} affects the derived upper limit on $\sum m_\nu$ from \ac{CMB} and \ac{BAO} data.

To quantitatively model the primordial perturbations, we adopt a general parameterization that includes both the adiabatic mode and the \ac{NDI} mode, following the formalism of Ref.~\cite{Bucher:2000kb}. Instead of assuming a power-law form for the power spectra, we treat the amplitudes at two distinct pivot scales as independent free parameters. Specifically, we choose \(k_1 = 0.002\,\text{Mpc}^{-1}\) and \(k_2 = 0.1\,\text{Mpc}^{-1}\), following Ref.~\cite{Planck:2018jri}. These two scales are selected to capture the complementary sensitivity of different cosmological probes: \(k_1\) lies in the regime of the \ac{CMB} Sachs--Wolfe plateau and the first acoustic peaks (\(\ell \sim 10\)--\(100\)), where \ac{NDI} modes leave their most prominent signature; \(k_2\) corresponds to smaller scales relevant for the \ac{CMB} damping tail (\(\ell \gtrsim 500\)) and the matter power spectrum wavenumbers probed by \ac{BAO} and weak lensing (\(k \sim 0.1\,h\,\text{Mpc}^{-1}\)), where the effects of neutrino free-streaming become significant. For each wavenumber \(k\), the primordial power spectrum is described by a real symmetric matrix
\(\mathcal{P}_{AB}(k)\) in the \((\mathcal{R},\mathcal{I})\) basis, where
\(A,B\in\{\mathcal{R},\mathcal{I}\}\) and the independent components are
\(\mathcal{P}_{\mathcal{R}\mathcal{R}}\), \(\mathcal{P}_{\mathcal{R}\mathcal{I}}\), and
\(\mathcal{P}_{\mathcal{I}\mathcal{I}}\) \cite{Planck:2018jri}.
where the diagonal entries \(\mathcal{P}_{\mathcal{R}\mathcal{R}}(k)\) and \(\mathcal{P}_{\mathcal{I}\mathcal{I}}(k)\) are interpolated between the two pivot scales using a piecewise constant or linear interpolation in \(\ln k\), and the off-diagonal term \(\mathcal{P}_{\mathcal{R}\mathcal{I}}(k)\) is parameterized by a correlation angle \(\Delta(k)\) such that \(\mathcal{P}_{\mathcal{R}\mathcal{I}}(k) = \sqrt{\mathcal{P}_{\mathcal{R}\mathcal{R}}(k) \mathcal{P}_{\mathcal{I}\mathcal{I}}(k)} \cos\Delta(k)\). The relative contribution of the NDI mode is conveniently expressed by the isocurvature fraction $\beta(k) = {\mathcal{P}_{\mathcal{I}\mathcal{I}}(k)}/(\mathcal{P}_{\mathcal{R}\mathcal{R}}(k) + \mathcal{P}_{\mathcal{I}\mathcal{I}}(k)),$
which may in general be scale-dependent. In this work, we consider the simple case of uncorrelated modes, i.e., \(\cos\Delta(k) = 0\) for all \(k\), which is a common assumption in the literature and is realized in many inflationary models with orthogonal field directions \cite{Langlois:1999dw, Gordon:2000hv}. The free parameters for the primordial power spectra are therefore the adiabatic amplitudes \(\mathcal{P}_{\mathcal{R}\mathcal{R}}^{(1)}\equiv \mathcal{P}_{\mathcal{R}\mathcal{R}}(k_1)\) and \(\mathcal{P}_{\mathcal{R}\mathcal{R}}^{(2)}\equiv \mathcal{P}_{\mathcal{R}\mathcal{R}}(k_2)\), and the \ac{NDI} amplitudes \(\mathcal{P}_{\mathcal{I}\mathcal{I}}^{(1)}\equiv \mathcal{P}_{\mathcal{I}\mathcal{I}}(k_1)\) and \(\mathcal{P}_{\mathcal{I}\mathcal{I}}^{(2)}\equiv \mathcal{P}_{\mathcal{I}\mathcal{I}}(k_2)\). 

The choice of piecewise interpolation (constant or linear) between the two pivots is a pragmatic simplification. We have not performed an explicit comparison of different interpolation schemes (e.g., log-linear vs. cubic spline) or tested the sensitivity of our results to the exact location of the pivot scales. A more flexible treatment, such as allowing the spectral indices of the \ac{NDI} and adiabatic modes as free parameters, would be desirable in future work but is beyond the scope of this paper. Nevertheless, given that the posterior constraints on \(\mathcal{P}_{\mathcal{I}\mathcal{I}}^{(1)}\) and \(\mathcal{P}_{\mathcal{I}\mathcal{I}}^{(2)}\) are both consistent with zero, our main conclusion about the robustness of \(\sum m_\nu\) limits against \ac{NDI} is unlikely to be significantly affected by these modeling choices.

The remaining cosmological parameters are the same as in the standard $\Lambda$CDM model: the physical baryon density $\omega_b = \Omega_b h^2$, the physical cold dark matter density $\omega_c = \Omega_c h^2$, the angular size of the sound horizon at recombination $\Theta_s$, the optical depth to reionization $\tau$, and the sum of neutrino masses $\sum m_\nu$. In total, our parameter space consists of $\{\omega_b, \omega_c, \Theta_s, \tau, \sum m_\nu\}$ plus the four amplitude parameters $\mathcal{P}_{\mathcal{R}\mathcal{R}}^{(1)}$, $\mathcal{P}_{\mathcal{R}\mathcal{R}}^{(2)}$, $\mathcal{P}_{\mathcal{I}\mathcal{I}}^{(1)}$, $\mathcal{P}_{\mathcal{I}\mathcal{I}}^{(2)}$. To study the robustness of neutrino mass constraints against the initial condition assumption, we compare two scenarios: (i) a pure adiabatic model where $\mathcal{P}_{\mathcal{I}\mathcal{I}}^{(1)}=\mathcal{P}_{\mathcal{I}\mathcal{I}}^{(2)}=0$; and (ii) a mixed model where these \ac{NDI} amplitudes are allowed to vary freely. In the mixed model, we adopt uniform priors on $\log \mathcal{P}_{\mathcal{R}\mathcal{R}} $ and $\log \mathcal{P}_{\mathcal{I}\mathcal{I}} $. All other parameters are assigned the same priors in both models to enable a fair assessment of how the inclusion of \ac{NDI} modes affects the upper limit on $\sum m_\nu$.

Recent cosmological observations, particularly from \ac{DESI} and \ac{DES}, have provided hints that dark energy may not be a strict cosmological constant but rather a dynamical component \cite{DESI:2025fii, DES:2025sig}. These analyses have shown mild tensions with the standard $\Lambda$CDM model and a preference for an evolving equation of state at low redshifts. To assess the robustness of our neutrino mass constraints in the context of dynamical dark energy, we also perform our analysis within the \ac{CPL} parameterization \cite{Chevallier:2000qy, Linder:2002et}. In the \ac{CPL} model, the dark energy equation of state is given by $w(a) = w_0 + w_a (1 - a) = w_0 + w_a {z}/{(1+z)}$, where $z$ stands for cosmological redshift defined as $z=1/a-1$ with $a$ being the cosmic scale, $w_0$ is the equation of state at the present day ($z=0$) and $w_a$ describes its redshift evolution. The background expansion history is modified accordingly, altering the angular diameter distance and the growth of structure, which in turn affects the \ac{CMB} and \ac{BAO} observables. In addition to the parameters already defined in our baseline $\Lambda$CDM model, the \ac{CPL} model introduces two independent parameters: $w_0$ and $w_a$. For the pure adiabatic and mixed  \ac{NDI} scenarios, we derive constraints on $\sum m_\nu$ both under the assumption of a cosmological constant ($w_0 = -1$, $w_a = 0$) and under the \ac{CPL} parameterization with flat priors $w_0 \in [-3, 1]$ and $w_a \in [-3, 2]$. This allows us to examine whether the robustness against initial conditions persists when the dark energy sector is allowed to vary.

In addition to $\sum m_\nu$, the neutrino mass hierarchy, i.e., the ordering of the three mass eigenstates, might affect cosmological observables. In the \ac{NH}, the three masses satisfy $m_1 < m_2 < m_3$; in the \ac{IH}, $m_3 < m_1 < m_2$. The difference manifests in the contribution of each eigenstate to the total $\sum m_\nu$ and in the free-streaming history, because the lightest eigenstate in \ac{IH} is heavier than in \ac{NH} for a fixed $\sum m_\nu$, altering the suppression scale of the matter power spectrum. However, given the sensitivity of current data, the impact of hierarchy is subtle and often degenerate with other parameters \cite{Lesgourgues:2006nd, Wong:2011ip, Huang:2015wrx, Wang:2017htc, Wang:2016tsz, Du:2025xes, Yang:2017amu,Guo:2018gyo,Du:2024pai}. To assess whether our conclusions on $\sum m_\nu$ depend on the assumed ordering, we perform our analysis under three scenarios: \ac{NH}, \ac{IH}, and a \ac{DH} where all three masses are set equal ($m_1=m_2=m_3 = \sum m_\nu/3$) with the same prior on $\sum m_\nu$ as in \ac{NH}/\ac{IH}. The \ac{DH} case serves as a reference for minimal hierarchy information. By comparing the posterior distributions and upper limits on $\sum m_\nu$ obtained from \ac{NH}, \ac{IH}, and \ac{DH}, we quantify the robustness of our neutrino mass constraints against the unknown mass ordering.

\section{Data and Method}\label{sec:data_method}

For our primary \ac{CMB} data, we adopt the \ac{CMB}-SPA likelihood \cite{SPT-3G:2025bzu}, which combines high-resolution temperature, polarization, and lensing measurements from Planck, \ac{ACT}, and \ac{SPTthreeG}. Specifically, we include the Planck 2018 low-$\ell$ TT and EE likelihoods for large angular scales, the \ac{ACT} \ac{DR6} primary \ac{CMB} likelihood for small-scale anisotropies, and the \ac{SPTthreeG} two-year delensed EE and lensing data to further sharpen the constraints on the damping tail and gravitational lensing potential \cite{Carron:2022eyg,SPT-3G:2024atg,SPT-3G:2025bzu,ACT:2025qjh,AtacamaCosmologyTelescope:2025blo}. To avoid double counting and reduce correlations between the Planck and \ac{ACT} datasets, we also use the Planck--\ac{ACT} cut combination that imposes appropriate multipole cuts, thereby ensuring a conservative and unbiased joint analysis \cite{AtacamaCosmologyTelescope:2025blo,AtacamaCosmologyTelescope:2025nti}.

\ac{BAO} measurements provide a robust standard ruler for probing the expansion history of the Universe. We use the most recent \ac{BAO} data from \ac{DESI} \ac{DR2}, which combines Lyman-$\alpha$ forest auto- and cross-correlations with galaxy and quasar clustering at lower redshifts \cite{DESI:2025zpo, DESI:2025zgx}. The \ac{DESI} \ac{DR2} sample is based on three years of observations, comprising over 14 million galaxies and quasars, and achieves a combined precision of $0.65\%$ on the isotropic \ac{BAO} scale at an effective redshift of $z_{\rm eff}=2.33$. The full \ac{DESI} \ac{DR2} \ac{BAO} likelihood covers a wide redshift range, providing a strong complement to \ac{CMB} data by breaking degeneracies between $\sum m_\nu$ and other parameters.

To further constrain the expansion history at moderate redshifts and to help break residual parameter degeneracies, we include the \ac{DES} Year 5 supernova sample \cite{DES:2024jxu, DES:2025sig}. This sample comprises approximately 1,500 spectroscopically confirmed and photometrically classified Type Ia \acp{SN} spanning redshifts $0.1 < z < 1.2$, and is the largest and deepest SN Ia sample to date. Including the \ac{SN} data helps to tighten constraints on $\Omega_m$ and the dark energy equation of state.

We perform parameter inference using MontePython \cite{Audren:2012wb, Brinckmann:2018cvx} interfaced with the \ac{CLASS} Boltzmann solver \cite{Blas:2011rf, Lesgourgues:2011rh}. For each cosmological model, \ac{CLASS} computes the theoretical \ac{CMB} temperature and polarization power spectra, the matter power spectrum, and the lensing potential, which are then passed to MontePython. MontePython implements a \ac{MCMC} sampler to explore the posterior distribution of the parameters. \ac{CLASS} is configured with high-precision settings to ensure accurate predictions up to the maximum multipoles of the \ac{CMB} data. For the neutrino sector, the  \ac{NDI} mode amplitude is implemented as an additional primordial power spectrum component within \ac{CLASS}.

\section{Results}\label{sec:results}

\begin{figure*}[htbp]
    \centering
    \includegraphics[width=0.7\linewidth]{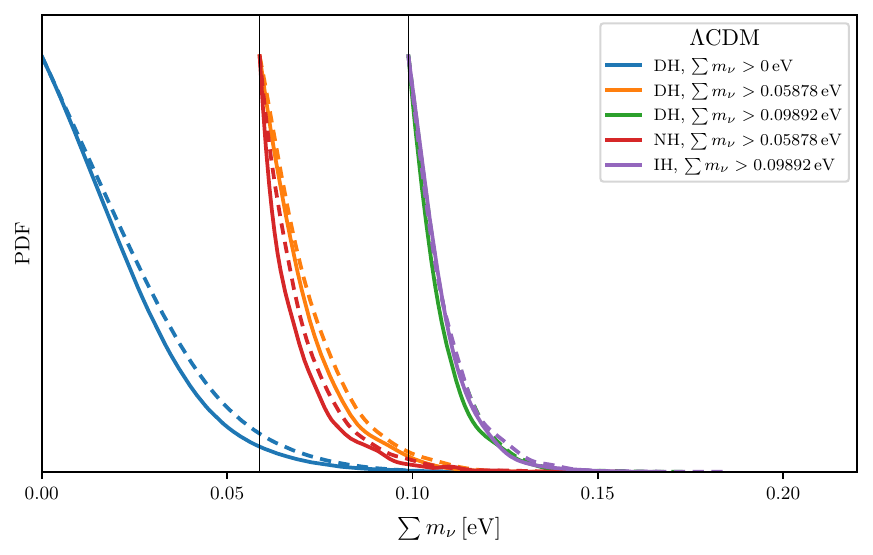}
    \caption{One-dimensional posterior distribution of $\sum m_\nu$ in the $\Lambda$CDM model under different initial condition assumptions. The solid lines represent the pure adiabatic initial conditions, while the dashed lines correspond to the mixed model including an \ac{NDI}component.}
    \label{fig:summnulcdm}
\end{figure*}

\begin{figure*}[htbp]
    \centering
    \includegraphics[width=0.7\linewidth]{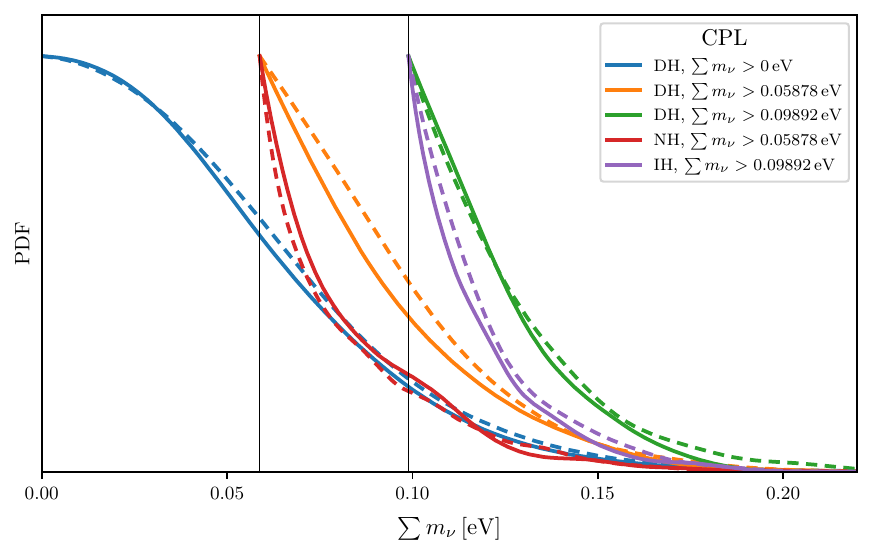}
    \caption{One-dimensional posterior distribution of $\sum m_\nu$ in the \ac{CPL} model under different initial condition assumptions. The solid lines represent the pure adiabatic initial conditions, while the dashed lines correspond to the mixed model including an \ac{NDI}component.}
    \label{fig:summnucpl}
\end{figure*}

We present our main results in terms of posterior distributions and summary statistics. For the standard $\Lambda$CDM model, Figure~\ref{fig:summnulcdm} shows the one-dimensional posterior distribution of $\sum m_\nu$, while the triangle plot of all cosmological parameters is provided in Appendix Figure~\ref{fig:lcdm_triangle}. The best-fit values, $1\sigma$ uncertainties, and the $95\%$ upper limit on $\sum m_\nu$ are summarized in Table~\ref{tab:lcdm_cpl}. For the \ac{CPL} dynamical dark energy model, the corresponding results, i.e., the $\sum m_\nu$ posterior, the triangle plot, and the summary statistics, are displayed in Figure~\ref{fig:summnucpl}, Appendix Figure~\ref{fig:cpl_triangle}, and Table~\ref{tab:lcdm_cpl}, respectively. In the following paragraphs, we describe the constraints obtained under each model assumption and discuss the robustness of the neutrino mass bound against the inclusion of  \ac{NDI}, dynamical dark energy, and the choice of neutrino mass hierarchy. Unless stated otherwise, all quoted limits assume the \ac{DH} and a uniform prior on $\sum m_\nu$ with lower bound equal to zero.

\begingroup
\begin{table*}[p]
\centering
\rotatebox{90}{%
\begin{minipage}{\textheight}
\centering
\caption{Marginalized constraints on cosmological parameters for the $\Lambda$CDM and \ac{CPL} models, considering different neutrino mass hierarchies (i.e., \ac{DH}, \ac{NH}, and \ac{IH}). All error bars represent the $68\%$ confidence interval, while the upper bounds for $\sum m_\nu$ are given at the $95\%$ confidence level.}
\label{tab:lcdm_cpl}
\resizebox{\textheight}{!}{%
\renewcommand{\arraystretch}{1.5}
\begin{tabular}{|c|l|c|c|c|c|c|c|c|c|c|c|}
\hline
Model & Params & \multicolumn{6}{c|}{DH} & \multicolumn{2}{c|}{NH} & \multicolumn{2}{c|}{IH} \\
\cline{3-12}
 & & \multicolumn{2}{c|}{$\sum m_\nu > 0\,{\rm eV}$} & \multicolumn{2}{c|}{$\sum m_\nu > 0.05878\,{\rm eV}$} & \multicolumn{2}{c|}{$\sum m_\nu > 0.09892\,{\rm eV}$} & \multicolumn{2}{c|}{$\sum m_\nu > 0.05878\,{\rm eV}$} & \multicolumn{2}{c|}{$\sum m_\nu > 0.09892\,{\rm eV}$} \\
\cline{3-12}
 & & AD & +NDI & AD & +NDI & AD & +NDI & AD & +NDI & AD & +NDI \\
\hline
\multirow{8}{*}{$\Lambda$CDM} & $\Omega_{\mathrm{b}} h^2$ & $0.02244\pm 0.00009$ & $0.02246\pm 0.00010$ & $0.02245\pm 0.00009$ & $0.02247\pm 0.00009$ & $0.02246\pm 0.00009$ & $0.02247\pm 0.00009$ & $0.02245\pm 0.00009$ & $0.02247\pm 0.00009$ & $0.02246\pm 0.00009$ & $0.02246\pm 0.00009$ \\
\cline{2-12}
 & $\Omega_{\mathrm{c}} h^2$ & $0.1187\pm 0.0006$ & $0.1188\pm 0.0006$ & $0.1184\pm 0.0006$ & $0.1185\pm 0.0006$ & $0.1182\pm 0.0006$ & $0.1183\pm 0.0006$ & $0.1184\pm 0.0006$ & $0.1186\pm 0.0006$ & $0.1182\pm 0.0006$ & $0.1183\pm 0.0006$ \\
\cline{2-12}
 & $100\theta_{\mathrm{s}}$ & $1.04176\pm 0.00022$ & $1.04202\pm 0.00028$ & $1.04179\pm 0.00023$ & $1.04207\pm 0.00028$ & $1.04185\pm 0.00023$ & $1.04214\pm 0.00029$ & $1.04179\pm 0.00023$ & $1.04210\pm 0.00028$ & $1.04182\pm 0.00022$ & $1.04216\pm 0.00029$ \\
\cline{2-12}
 & $\tau$ & $0.0576\pm 0.0039$ & $0.0573\pm 0.0039$ & $0.0592\pm 0.0039$ & $0.0586\pm 0.0039$ & $0.0610\pm 0.0039$ & $0.0603\pm 0.0040$ & $0.0596\pm 0.0040$ & $0.0586\pm 0.0040$ & $0.0612\pm 0.0041$ & $0.0603\pm 0.0040$ \\
\cline{2-12}
 & $10^9\mathcal{P}_{\mathcal{R}\mathcal{R}}^{(1)}$ & $2.314\pm 0.030$ & $2.311\pm 0.034$ & $2.317\pm 0.030$ & $2.318\pm 0.036$ & $2.323\pm 0.030$ & $2.323\pm 0.035$ & $2.320\pm 0.030$ & $2.320\pm 0.036$ & $2.324\pm 0.030$ & $2.326\pm 0.036$ \\
\cline{2-12}
 & $10^9\mathcal{P}_{\mathcal{R}\mathcal{R}}^{(2)}$ & $2.073\pm 0.016$ & $2.060\pm 0.019$ & $2.082\pm 0.017$ & $2.067\pm 0.019$ & $2.092\pm 0.017$ & $2.075\pm 0.020$ & $2.084\pm 0.018$ & $2.067\pm 0.020$ & $2.092\pm 0.017$ & $2.075\pm 0.020$ \\
\cline{2-12}
 & $10^{10}\mathcal{P}_{\mathcal{I}\mathcal{I}}^{(1)}$ & / & $0.57^{+0.57}_{-0.49}$ & / & $0.52\pm 0.47$ & / & $0.58^{+0.59}_{-0.50}$ & / & $0.48\pm 0.43$ & / & $0.48\pm 0.43$ \\
\cline{2-12}
 & $10^{10}\mathcal{P}_{\mathcal{I}\mathcal{I}}^{(2)}$ & / & $1.9\pm 1.3$ & / & $2.2\pm 1.4$ & / & $2.3\pm 1.5$ & / & $2.3\pm 1.5$ & / & $2.4\pm 1.6$ \\
\cline{2-12}
 & $\sum m_\nu$ & $< 0.052\,{\rm eV}$ & $< 0.057\,{\rm eV}$ & $< 0.092\,{\rm eV}$ & $< 0.094\,{\rm eV}$ & $< 0.123\,{\rm eV}$ & $< 0.124\,{\rm eV}$ & $< 0.092\,{\rm eV}$ & $< 0.092\,{\rm eV}$ & $< 0.123\,{\rm eV}$ & $< 0.126\,{\rm eV}$ \\
\hline
\hline
\multirow{10}{*}{CPL} & $\Omega_{\mathrm{b}} h^2$ & $0.02242\pm 0.00009$ & $0.02243\pm 0.00010$ & $0.02241\pm 0.00009$ & $0.02244\pm 0.00009$ & $0.02242\pm 0.00009$ & $0.02243\pm 0.00010$ & $0.02242\pm 0.00009$ & $0.02242\pm 0.00009$ & $0.02241\pm 0.00009$ & $0.02243\pm 0.00010$ \\
\cline{2-12}
 & $\Omega_{\mathrm{c}} h^2$ & $0.1195\pm 0.0007$ & $0.1195\pm 0.0007$ & $0.1196\pm 0.0006$ & $0.1196\pm 0.0007$ & $0.1196\pm 0.0006$ & $0.1196\pm 0.0007$ & $0.1196\pm 0.0007$ & $0.1196\pm 0.0007$ & $0.1196\pm 0.0007$ & $0.1196\pm 0.0007$ \\
\cline{2-12}
 & $100\theta_{\mathrm{s}}$ & $1.04168\pm 0.00023$ & $1.04191\pm 0.00027$ & $1.04169\pm 0.00023$ & $1.04194\pm 0.00028$ & $1.04168\pm 0.00023$ & $1.04196\pm 0.00030$ & $1.04170\pm 0.00022$ & $1.04193\pm 0.00028$ & $1.04170\pm 0.00023$ & $1.04194\pm 0.00028$ \\
\cline{2-12}
 & $\tau$ & $0.0557\pm 0.0039$ & $0.0555\pm 0.0039$ & $0.0562\pm 0.0038$ & $0.0559\pm 0.0039$ & $0.0569\pm 0.0037$ & $0.0567\pm 0.0040$ & $0.0564\pm 0.0040$ & $0.0559\pm 0.0039$ & $0.0569\pm 0.0042$ & $0.0565\pm 0.0039$ \\
\cline{2-12}
 & $w_0$ & $-0.811\pm 0.056$ & $-0.810\pm 0.058$ & $-0.791\pm 0.055$ & $-0.792\pm 0.058$ & $-0.772\pm 0.055$ & $-0.777\pm 0.056$ & $-0.792\pm 0.054$ & $-0.792\pm 0.054$ & $-0.772\pm 0.053$ & $-0.779\pm 0.056$ \\
\cline{2-12}
 & $w_a$ & $-0.70\pm 0.22$ & $-0.68\pm 0.23$ & $-0.83\pm 0.22$ & $-0.82\pm 0.23$ & $-0.94\pm 0.22$ & $-0.92\pm 0.23$ & $-0.82\pm 0.21$ & $-0.81\pm 0.22$ & $-0.94\pm 0.22$ & $-0.91\pm 0.22$ \\
\cline{2-12}
 & $10^9\mathcal{P}_{\mathcal{R}\mathcal{R}}^{(1)}$ & $2.318\pm 0.028$ & $2.314\pm 0.035$ & $2.320\pm 0.030$ & $2.322\pm 0.034$ & $2.328\pm 0.029$ & $2.326\pm 0.034$ & $2.323\pm 0.030$ & $2.320\pm 0.034$ & $2.328\pm 0.029$ & $2.325\pm 0.034$ \\
\cline{2-12}
 & $10^9\mathcal{P}_{\mathcal{R}\mathcal{R}}^{(2)}$ & $2.062\pm 0.017$ & $2.052\pm 0.019$ & $2.065\pm 0.017$ & $2.054\pm 0.019$ & $2.070\pm 0.016$ & $2.059\pm 0.020$ & $2.066\pm 0.017$ & $2.055\pm 0.020$ & $2.070\pm 0.018$ & $2.059\pm 0.019$ \\
\cline{2-12}
 & $10^{10}\mathcal{P}_{\mathcal{I}\mathcal{I}}^{(1)}$ & / & $0.60^{+0.58}_{-0.52}$ & / & $0.57\pm 0.52$ & / & $0.60\pm 0.54$ & / & $0.58^{+0.59}_{-0.52}$ & / & $0.55^{+0.54}_{-0.48}$ \\
\cline{2-12}
 & $10^{10}\mathcal{P}_{\mathcal{I}\mathcal{I}}^{(2)}$ & / & $1.6\pm 1.2$ & / & $1.8\pm 1.3$ & / & $1.9\pm 1.3$ & / & $1.7\pm 1.3$ & / & $1.8\pm 1.3$ \\
\cline{2-12}
 & $\sum m_\nu$ & $< 0.111\,{\rm eV}$ & $< 0.115\,{\rm eV}$ & $< 0.142\,{\rm eV}$ & $< 0.142\,{\rm eV}$ & $< 0.158\,{\rm eV}$ & $< 0.173\,{\rm eV}$ & $< 0.130\,{\rm eV}$ & $< 0.136\,{\rm eV}$ & $< 0.154\,{\rm eV}$ & $< 0.156\,{\rm eV}$ \\
\hline

\end{tabular}%
}
\end{minipage}
}
\end{table*}
\endgroup

Before proceeding to the constraints, we highlight the distinct scale-dependent signatures of massive neutrinos and \ac{NDI} modes in cosmological observables, which underpin their distinguishability in current data. As illustrated in the left panels of Figs.~\ref{fig:cmb_ndi} and~\ref{fig:bao_ndi}, increasing $\sum m_\nu$ (under fixed adiabatic initial conditions) primarily suppresses the \ac{CMB} temperature and polarization power spectra on small angular scales ($\ell \gtrsim 500$), i.e., the damping tail, and suppresses the matter power spectrum $P_m(k)$ over a broad range of wavenumbers $k \gtrsim 0.01\,h\,\text{Mpc}^{-1}$ due to free-streaming. In contrast, the \ac{NDI} mode (right panels) modifies the \ac{CMB} power spectra on larger angular scales ($\ell \sim 10$--$100$), altering the relative heights of the first few acoustic peaks while leaving the small-scale damping tail largely unchanged. In the matter power spectrum, the \ac{NDI} component induces a scale-dependent suppression that peaks at intermediate wavenumbers $k \sim 0.01$--$0.1\,h\,\text{Mpc}^{-1}$, a direct consequence of the absence of an initial curvature perturbation. This clear separation of scales explains why current high-precision data (e.g., \ac{ACT} \ac{DR6} and \ac{SPTthreeG} for small-scale \ac{CMB}, and \ac{DESI} \ac{DR2} for intermediate-scale structure) can break the degeneracy between $\sum m_\nu$ and the \ac{NDI} amplitude. Consequently, the fact that our posterior for the \ac{NDI} amplitude is consistent with zero (Sec.~\ref{sec:results}) is not only a lack of sensitivity, but also a positive indication that the data favor purely adiabatic initial conditions, with no evidence for a sizeable isocurvature component.

First, we consider the standard $\Lambda$CDM cosmology. Under purely adiabatic initial conditions, we obtain a $95\%$ upper limit of 
\begin{equation}
\sum m_\nu < 0.052 \ \mathrm{eV}\,.\ (\mathrm{DH,\ adiabatic})
\end{equation}
When we allow for an additional  \ac{NDI} component, the bound shifts only marginally to 
\begin{equation}
\sum m_\nu < 0.057 \ \mathrm{eV}\,.\ (\mathrm{DH,\ isocurvature})
\end{equation}
The  \ac{NDI} amplitude itself is tightly constrained: the $95\%$ upper limit on the isocurvature power amplitude is consistent with zero, indicating no evidence for a non-adiabatic component. Thus, the inclusion of isocurvature modes does not significantly weaken the neutrino mass limit within $\Lambda$CDM, consistent with the lack of evidence for sizeable \ac{NDI} contributions in current CMB and large-scale-structure data. Turning to the mass hierarchy, we compare \ac{NH}, \ac{IH}, and \ac{DH}. The $95\%$ upper limits for \ac{NH} and \ac{IH} differ from the \ac{DH} result by less than $0.002$ eV when using the same prior lower bound. Importantly, the lower bound of the prior on $\sum m_\nu$ has a noticeable impact: enforcing a positive minimum (e.g., $\sum m_\nu \ge 0.06$ eV) shifts the posterior upward, raising the $95\%$ upper bound by $\sim0.03$--$0.04$ eV relative to the baseline prior $\sum m_\nu \in [0,3]$ eV. This sensitivity arises because the likelihood permits very small $\sum m_\nu$; truncating that region forces higher masses.

We now turn to the \ac{CPL} dynamical dark energy model. Under adiabatic-only assumptions, the $95\%$ upper limit on $\sum m_\nu$ loosens
\begin{equation}
\sum m_\nu < 0.111 \ \mathrm{eV}\,,\ (\mathrm{DH,\ adiabatic})
\end{equation}
reflecting degeneracies with $(w_0,w_a)$. Adding an  \ac{NDI} component further relaxes the bound to
\begin{equation}
\sum m_\nu < 0.115 \ \mathrm{eV}\,,\ (\mathrm{DH,\ isocurvature})
\end{equation}
yet the  \ac{NDI} amplitude remains consistent with zero, indicating no statistically meaningful isocurvature contribution. The dark energy parameters are constrained to $w_0 = -0.811{\pm0.056}$ and $w_a = -0.70{\pm0.22}$ for the adiabatic case, with the mild preference for $w_a<0$ seen in recent \ac{DESI}+\ac{DES} analyses persisting in the adiabatic scenario. When  \ac{NDI} is included, the central values shift by less than $0.02$ for both $w_0$ and $w_a$, well within the $1\sigma$ errors. Regarding the mass hierarchy, the differences among \ac{NH}, \ac{IH}, and \ac{DH} are again minimal: the $95\%$ upper limits vary by at most $0.017$ eV when using the same prior lower bound, and the prior lower bound also influences the limit here; setting $\sum m_\nu \ge 0.05878$ eV increases the upper bound by about $0.03$ eV, similar to the $\Lambda$CDM case.

Comparing the two dark energy models, the impact on the neutrino mass bound is substantial: the $\sum m_\nu$ upper limit increases from $0.052$ eV ($\Lambda$CDM, adiabatic, \ac{DH}, prior lower bound $0$) to $0.111$ eV (\ac{CPL}, adiabatic, \ac{DH}, prior lower bound $0$), leading to a $50\%$ weakening. This demonstrates that the inferred mass scale is sensitive to assumptions about the late-time expansion history, because a larger $\sum m_\nu$ suppresses structure growth, an effect that can be partially compensated by an evolving dark energy equation of state. Nevertheless, the robustness against  \ac{NDI} holds in both models: adding isocurvature changes the bound by at most $0.015$ eV, comparable to the statistical uncertainty.

It is useful to quantify the improvement relative to earlier data combinations. The Planck 2018 result gave $\sum m_\nu<0.12,{\rm eV}$ at $95\%$ confidence when combined with previously-released \ac{BAO} data \cite{Planck:2018vyg}. Our baseline $\Lambda$CDM adiabatic result from the full \ac{CMB}-SPA+\ac{DESI} \ac{DR2}+\ac{DES} \acp{SN} combination, $\sum m_\nu<0.052,{\rm eV}$, is therefore tighter by a factor of $0.12/0.052\simeq2.3$. Even after allowing for an \ac{NDI} component, the corresponding bound, $\sum m_\nu<0.057,{\rm eV}$, remains tighter by a factor of $0.12/0.057\simeq2.1$. For the isocurvature sector, we find $10^{10}\mathcal{P}_{\mathcal{I}\mathcal{I}}^{(1)}=0.57^{+0.57}_{-0.49}$ and $10^{10}\mathcal{P}_{\mathcal{I}\mathcal{I}}^{(2)}=1.9\pm1.3$ in our baseline $\Lambda$CDM+\ac{NDI} case. These constraints are broadly comparable to the Planck 2018 uncorrelated-\ac{NDI} constraints at the large-scale pivot and are tighter at the small-scale pivot when compared with the free-tilt Planck result \cite{Planck:2018jri}. Thus, while we do not claim a uniform improvement for every \ac{NDI} amplitude parameter, the present high-resolution \ac{CMB} and \ac{BAO} data substantially sharpen the neutrino-mass bound and improve the small-scale \ac{NDI} constraint.

Overall, this work provides the first joint constraint on $\sum m_\nu$ and  \ac{NDI} using the full \ac{CMB}-SPA+\ac{DESI} \ac{DR2} \ac{BAO}+\ac{DES} \acp{SN} dataset, demonstrating that current data are insensitive to neutrino isocurvature modes and that the neutrino mass upper limit is robust against this extension, though it strongly depends on the assumption of dark energy.

\section{Conclusions and Discussion}\label{sec:conclusion}

In this work, we have investigated the robustness of cosmological upper limits on the sum of neutrino masses against the assumption of primordial initial conditions. Specifically, we considered the possibility of an  \ac{NDI} component in addition to the standard adiabatic mode, and we analyzed the latest observational data sets: the \ac{CMB}-SPA combination (Planck 2018, \ac{ACT} \ac{DR6}, and \ac{SPTthreeG}), \ac{DESI} \ac{DR2} \ac{BAO}, and the \ac{DES} Year 5 supernova sample. Our analysis was performed within both the standard $\Lambda$CDM framework and the \ac{CPL} dynamical dark energy model, and we also examined the impact of the neutrino mass hierarchy (\ac{NH}, \ac{IH}, and \ac{DH}) as well as the effect of the prior lower bound on $\sum m_\nu$.

Our main findings are as follows. First, within $\Lambda$CDM, the $95\%$ upper limit on $\sum m_\nu$ under purely adiabatic initial conditions is $0.052$ eV (assuming \ac{DH} and a prior lower bound of $0$ eV). When an  \ac{NDI} component is added, the bound increases only marginally to $0.057$ eV, and the  \ac{NDI} amplitude is consistent with zero at $95\%$ confidence. Hence, the neutrino mass constraint is robust against the inclusion of isocurvature modes in a $\Lambda$CDM universe. Second, in the \ac{CPL} model, the adiabatic upper limit loosens to $0.111$ eV, and adding  \ac{NDI} further relaxes it to $0.115$ eV, indicating a $50\%$ increase relative to $\Lambda$CDM. This demonstrates that the inferred neutrino mass scale is highly sensitive to assumptions about the dark energy equation of state. Nevertheless, even in the \ac{CPL} model, the  \ac{NDI} amplitude remains consistent with zero, and the isocurvature component does not significantly alter the $\sum m_\nu$ bound. Third, we studied the influence of the neutrino mass hierarchy. Current cosmological data cannot distinguish the mass ordering. However, the prior lower bound on $\sum m_\nu$ has a notable effect: enforcing a positive minimum (e.g., $0.05878$ eV) raises the upper limit by $0.03$–$0.04$ eV. Therefore, when quoting upper limits, it is essential to specify the prior range, especially the lower bound.

The implications of our results are twofold. On the one hand, the robustness of the $\sum m_\nu$ upper limit against  \ac{NDI} modes suggests that current cosmological bounds are reliable even if the primordial perturbations are not purely adiabatic, provided that the  \ac{NDI} amplitude is small. On the other hand, the strong dependence on the dark energy model underscores the need for a precise determination of the late-time expansion history before a definitive, model-independent neutrino mass limit can be claimed.

Our finding that \ac{NDI} modes do not significantly weaken the $\sum m_\nu$ bound raises the question: is this robustness generic, or does it arise because the current limit lies close to the minimal mass allowed by the neutrino hierarchy? The minimum $\sum m_\nu$ is $0.05878$ eV for \ac{NH} and $0.09892$ eV for \ac{IH}. Our $\Lambda$CDM adiabatic upper limit ($0.052$ eV) sits just below the \ac{NH} floor; imposing a physically motivated lower bound (e.g., $0.05878$ eV) raises the limit to $0.092$ eV, while the \ac{NDI}$-$allowed bound becomes $0.094$ eV, indicating an increase of only $0.002$ eV. This tiny shift reveals that the data already constrain $\sum m_\nu$ to a narrow window above the hierarchy minimum. In this regime, the degeneracy between $\sum m_\nu$ and the \ac{NDI} amplitude is limited: a larger $\sum m_\nu$ would over‑suppress structure, and the distinct scale‑dependent signature of \ac{NDI} cannot fully compensate. Hence, even allowing \ac{NDI} hardly relaxes the bound. The same holds for the \ac{CPL} model, where the absolute limits are higher ($0.111$–$0.115$ eV) but the relative weakening due to \ac{NDI} remains minimal ($0.004$ eV). 

Physically motivated priors should at least incorporate the minimum mass implied by the neutrino mass hierarchy, i.e., 0.05878 eV for \ac{NH} and 0.09892 eV for \ac{IH}. Our adiabatic $\Lambda$CDM upper limit of $0.052$ eV lies below the \ac{NH} minimum, indicating that this bound is an artifact of the statistical prior extending to zero rather than a physically allowed mass range. Consequently, we recommend reporting two limits simultaneously: one with a zero lower bound (to facilitate comparison with previous work) and one with a hierarchy-informed lower bound (to reflect physical plausibility). For instance, in $\Lambda$CDM with adiabatic initial conditions, the $95\%$ upper limit becomes $<0.092$ eV when imposing $\sum m_\nu \ge 0.05878$ eV, compared to $<0.052$ eV under the zero lower bound. In summary, the observed robustness is intimately tied to the fact that current data already confine $\sum m_\nu$ to a narrow range just above the minimal hierarchical mass, leaving little room for isocurvature modes to alter the limit, but the absolute value of the upper limit is strongly prior-dependent and should be interpreted with the hierarchy minimum in mind.

Although our analysis shows that current data disfavor a significant \ac{NDI} component, other new physics can mimic or degenerate with the effects of massive neutrinos. Dark radiation with self-interactions, for instance, can produce scale-dependent signatures in the \ac{CMB} and matter power spectra that partially resemble those of \ac{NDI} modes \cite{Chang:2025uvx}. When the neutrino energy density is allowed to vary, cosmological data prefer massless free-streaming dark radiation over massive neutrinos, significantly relaxing the neutrino mass bound \cite{Chang:2025uvx}. Moreover, neutrino self-interactions beyond the Standard Model have profound cosmological implications and can alter \ac{CMB} anisotropies in ways that may degenerate with \ac{NDI} signatures \cite{Berryman:2022hds}. Given that our constraints on the \ac{NDI} amplitude are consistent with zero, a joint analysis combining \ac{NDI} with extended neutrino sectors (e.g., $N_{\mathrm{eff}}$ or self-interactions) would be a natural next step.

Next-generation surveys will dramatically sharpen constraints on both $\sum m_\nu$ and \ac{NDI} modes. The combination of \ac{CMB}-S4 and LiteBIRD is forecast to reduce isocurvature error bars by up to $70\%$ for neutrino isocurvature modes \cite{Montandon:2020kuk}. In the large-scale-structure sector, the Euclid alone is expected to achieve $\sigma(\sum m_\nu) \approx 56$ meV in $\Lambda$CDM, and combined with Planck this improves to $\approx 23$ meV, offering evidence of a non-zero neutrino mass to at least $2.6\sigma$ \cite{Euclid:2024imf}. Including future \ac{CMB} data from LiteBIRD and \ac{CMB}-S4 could push this to a $4\sigma$ detection \cite{Euclid:2024imf}. Should a non-zero \ac{NDI} component exist, these future surveys would have the sensitivity to detect it; otherwise, they will firmly establish the adiabatic nature of primordial perturbations.

\acknowledgments

S.W. is supported by the National Natural Science Foundation of China (Grant No. 12533001). 
Z.C.Z. is supported by the National Key Research and Development Program of China (Grant No. 2021YFC2203001). 
X.Z. is supported by the National Natural Science Foundation of China (Grants Nos. 12473001, 12575049, 12533001), the National SKA Program of China (Grants Nos. 2022SKA0110200, 2022SKA0110203), and the China Manned Space Program (Grant No. CMS-CSST-2025-A02). 
This study was supported by the Advanced Computation Center of Hangzhou Normal University and the High-performance Computing Platform of China Agricultural University.

\appendix

\begin{widetext}

\section{Results from Parameter Inference}

The triangle plots of all cosmological parameters for the $\Lambda$CDM and \ac{CPL} models are provided in Figures~\ref{fig:lcdm_triangle} and~\ref{fig:cpl_triangle}, respectively.

\begin{figure}[htbp]
    \centering
    \includegraphics[width=1\textwidth]{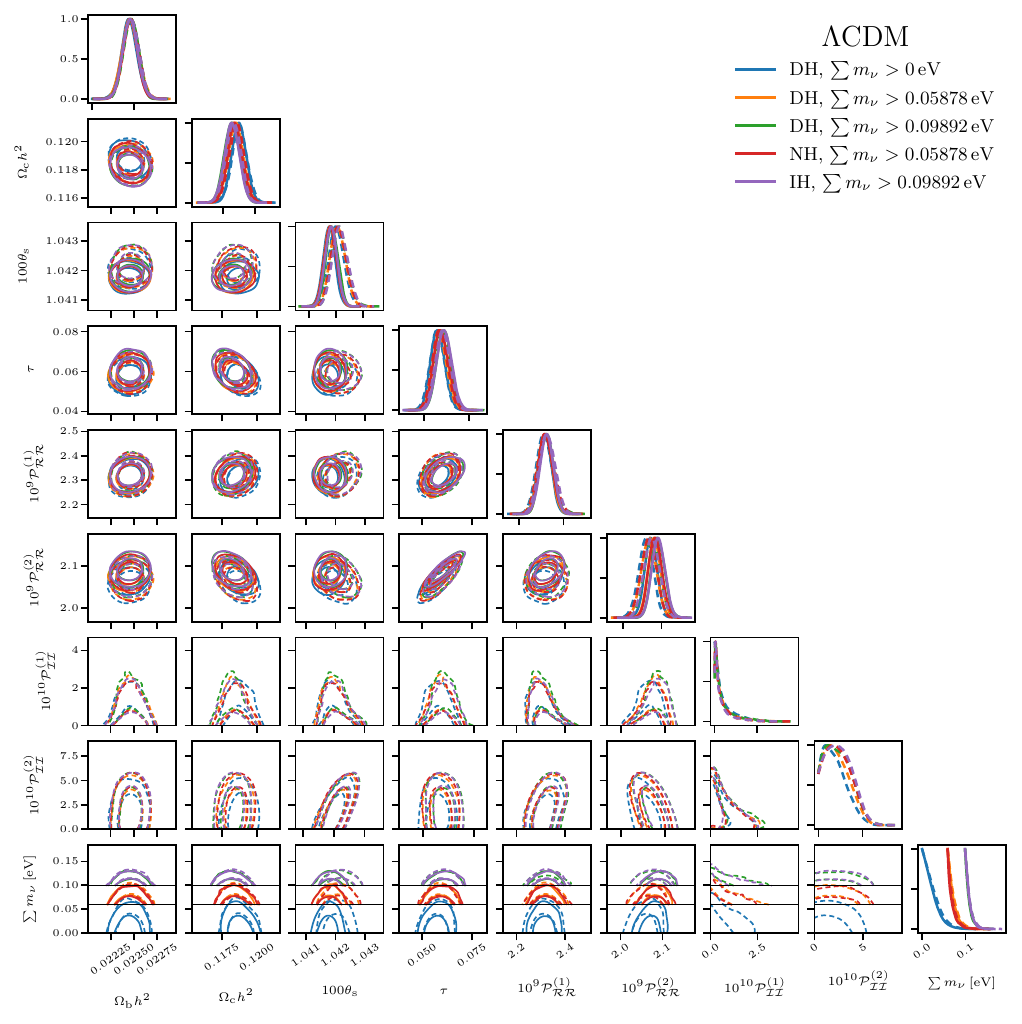}
    \caption{Triangle plot of all cosmological parameters in the $\Lambda$CDM model. The solid lines represent the pure adiabatic initial conditions, while the dashed lines correspond to the mixed model including an \ac{NDI}component.}
    \label{fig:lcdm_triangle}
\end{figure}

\begin{figure}[htbp]
    \centering
    \includegraphics[width=1\textwidth]{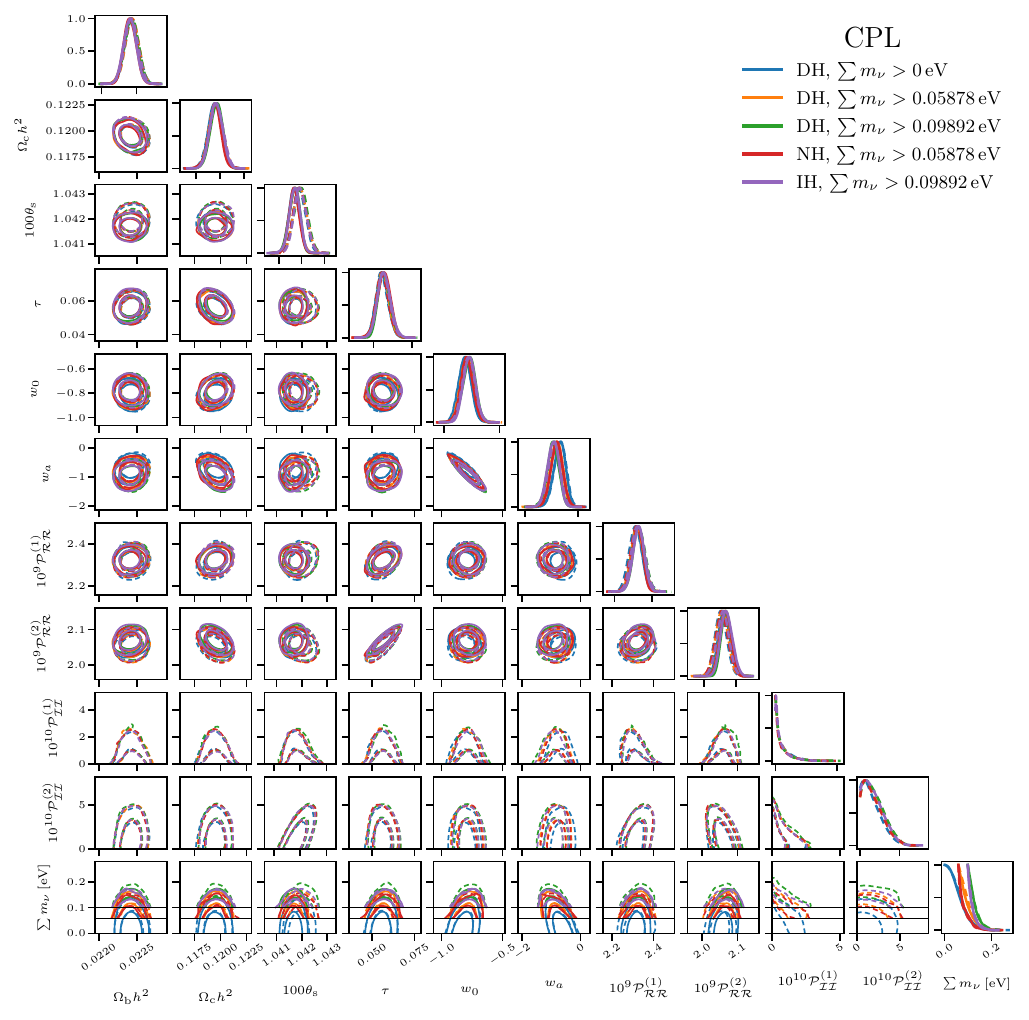}
    \caption{Triangle plot of all cosmological parameters in the \ac{CPL} model. The solid lines represent the pure adiabatic initial conditions, while the dashed lines correspond to the mixed model including an \ac{NDI}component.}
    \label{fig:cpl_triangle}
\end{figure}

\end{widetext}

\bibliographystyle{apsrev4-1}
\bibliography{biblio}

\end{document}